\RequirePackage{lineno}

\documentclass[aps,prl,twocolumn,superscriptaddress,showpacs,preprintnumbers,amsmath,amssymb]{revtex4-1}

\usepackage{graphicx} 
\usepackage{dcolumn}  
\usepackage{color} %
\RequirePackage{xspace}
\usepackage{relsize}

\graphicspath{{ps}}

\def\Ps      {\ensuremath{s}\xspace}  
\def\squark    {\ensuremath{\Ps}\xspace}
 \def\PK      {\ensuremath{K}\xspace}  
\def\kaon  {\ensuremath{\PK}\xspace}

\def\Kzb   {\overline{K}{}^{\,0}}

\def\Kz    {\ensuremath{\kaon^0}\xspace}
\def\KS    {\ensuremath{\kaon^0_{S}}\xspace}

\def\Ppi         {\ensuremath{\pi}\xspace}   
\def\pion  {\ensuremath{\Ppi}\xspace}
\def\pip   {\ensuremath{\pion^+}\xspace}
\def\pim   {\ensuremath{\pion^-}\xspace}

\def\to       {\ensuremath{\rightarrow}\xspace} 
\def\qbar     {\kern 0.2em\overline{\kern -0.2em q}{}\xspace}
\def\evtgen   {\mbox{\textsc{EvtGen}}\xspace}
\def\geant    {\mbox{\textsc{Geant3}}\xspace}

\def\infb     {fb$^{-1}$}

\def\Bst      {\ensuremath{B^{*0}_\squark}\xspace}
\def\Bstar      {\ensuremath{B^{(*)0}_\squark}\xspace}
\def\Bs{B^0_s}
\def\Bsb{\overline{B}{}^{\,0}_s}

\def\Bstb    {\overline{B}{}^{\,*0}_s}
\def\Bstarb  {\overline{B}{}^{\,(*)0}_s}

\def\mevm     {MeV/$c^2$}

\def\gevm     {GeV/$c^2$}

\setpagewiselinenumbers
\begin{document}


\vskip -1.0cm
\title{ \quad\\[1.0cm] 
Observation of the decay \mbox{\boldmath$\Bs\to\Kz\Kzb$} }
\noaffiliation
\affiliation{Aligarh Muslim University, Aligarh 202002}
\affiliation{University of the Basque Country UPV/EHU, 48080 Bilbao}
\affiliation{University of Bonn, 53115 Bonn}
\affiliation{Budker Institute of Nuclear Physics SB RAS, Novosibirsk 630090}
\affiliation{Faculty of Mathematics and Physics, Charles University, 121 16 Prague}
\affiliation{Chonnam National University, Kwangju 660-701}
\affiliation{University of Cincinnati, Cincinnati, Ohio 45221}
\affiliation{Deutsches Elektronen--Synchrotron, 22607 Hamburg}
\affiliation{University of Florida, Gainesville, Florida 32611}
\affiliation{Justus-Liebig-Universit\"at Gie\ss{}en, 35392 Gie\ss{}en}
\affiliation{Gifu University, Gifu 501-1193}
\affiliation{SOKENDAI (The Graduate University for Advanced Studies), Hayama 240-0193}
\affiliation{Hanyang University, Seoul 133-791}
\affiliation{University of Hawaii, Honolulu, Hawaii 96822}
\affiliation{High Energy Accelerator Research Organization (KEK), Tsukuba 305-0801}
\affiliation{IKERBASQUE, Basque Foundation for Science, 48013 Bilbao}
\affiliation{Indian Institute of Technology Bhubaneswar, Satya Nagar 751007}
\affiliation{Indian Institute of Technology Guwahati, Assam 781039}
\affiliation{Indian Institute of Technology Madras, Chennai 600036}
\affiliation{Indiana University, Bloomington, Indiana 47408}
\affiliation{Institute of High Energy Physics, Chinese Academy of Sciences, Beijing 100049}
\affiliation{Institute of High Energy Physics, Vienna 1050}
\affiliation{Institute for High Energy Physics, Protvino 142281}
\affiliation{INFN - Sezione di Torino, 10125 Torino}
\affiliation{Institute for Theoretical and Experimental Physics, Moscow 117218}
\affiliation{J. Stefan Institute, 1000 Ljubljana}
\affiliation{Kanagawa University, Yokohama 221-8686}
\affiliation{Institut f\"ur Experimentelle Kernphysik, Karlsruher Institut f\"ur Technologie, 76131 Karlsruhe}
\affiliation{Kennesaw State University, Kennesaw GA 30144}
\affiliation{King Abdulaziz City for Science and Technology, Riyadh 11442}
\affiliation{Korea Institute of Science and Technology Information, Daejeon 305-806}
\affiliation{Korea University, Seoul 136-713}
\affiliation{Kyungpook National University, Daegu 702-701}
\affiliation{\'Ecole Polytechnique F\'ed\'erale de Lausanne (EPFL), Lausanne 1015}
\affiliation{Faculty of Mathematics and Physics, University of Ljubljana, 1000 Ljubljana}
\affiliation{Ludwig Maximilians University, 80539 Munich}
\affiliation{Luther College, Decorah, Iowa 52101}
\affiliation{University of Maribor, 2000 Maribor}
\affiliation{Max-Planck-Institut f\"ur Physik, 80805 M\"unchen}
\affiliation{School of Physics, University of Melbourne, Victoria 3010}
\affiliation{Moscow Physical Engineering Institute, Moscow 115409}
\affiliation{Moscow Institute of Physics and Technology, Moscow Region 141700}
\affiliation{Graduate School of Science, Nagoya University, Nagoya 464-8602}
\affiliation{Kobayashi-Maskawa Institute, Nagoya University, Nagoya 464-8602}
\affiliation{Nara Women's University, Nara 630-8506}
\affiliation{National Central University, Chung-li 32054}
\affiliation{National United University, Miao Li 36003}
\affiliation{Department of Physics, National Taiwan University, Taipei 10617}
\affiliation{H. Niewodniczanski Institute of Nuclear Physics, Krakow 31-342}
\affiliation{Niigata University, Niigata 950-2181}
\affiliation{University of Nova Gorica, 5000 Nova Gorica}
\affiliation{Novosibirsk State University, Novosibirsk 630090}
\affiliation{Osaka City University, Osaka 558-8585}
\affiliation{Pacific Northwest National Laboratory, Richland, Washington 99352}
\affiliation{Peking University, Beijing 100871}
\affiliation{University of Pittsburgh, Pittsburgh, Pennsylvania 15260}
\affiliation{Punjab Agricultural University, Ludhiana 141004}
\affiliation{University of Science and Technology of China, Hefei 230026}
\affiliation{Seoul National University, Seoul 151-742}
\affiliation{Soongsil University, Seoul 156-743}
\affiliation{University of South Carolina, Columbia, South Carolina 29208}
\affiliation{Sungkyunkwan University, Suwon 440-746}
\affiliation{School of Physics, University of Sydney, NSW 2006}
\affiliation{Department of Physics, Faculty of Science, University of Tabuk, Tabuk 71451}
\affiliation{Tata Institute of Fundamental Research, Mumbai 400005}
\affiliation{Excellence Cluster Universe, Technische Universit\"at M\"unchen, 85748 Garching}
\affiliation{Department of Physics, Technische Universit\"at M\"unchen, 85748 Garching}
\affiliation{Toho University, Funabashi 274-8510}
\affiliation{Department of Physics, Tohoku University, Sendai 980-8578}
\affiliation{Earthquake Research Institute, University of Tokyo, Tokyo 113-0032}
\affiliation{Department of Physics, University of Tokyo, Tokyo 113-0033}
\affiliation{Tokyo Institute of Technology, Tokyo 152-8550}
\affiliation{Tokyo Metropolitan University, Tokyo 192-0397}
\affiliation{University of Torino, 10124 Torino}
\affiliation{Utkal University, Bhubaneswar 751004}
\affiliation{CNP, Virginia Polytechnic Institute and State University, Blacksburg, Virginia 24061}
\affiliation{Wayne State University, Detroit, Michigan 48202}
\affiliation{Yamagata University, Yamagata 990-8560}
\affiliation{Yonsei University, Seoul 120-749}
  \author{B.~Pal}\affiliation{University of Cincinnati, Cincinnati, Ohio 45221} 
  \author{A.~J.~Schwartz}\affiliation{University of Cincinnati, Cincinnati, Ohio 45221} 
  \author{A.~Abdesselam}\affiliation{Department of Physics, Faculty of Science, University of Tabuk, Tabuk 71451} 
  \author{I.~Adachi}\affiliation{High Energy Accelerator Research Organization (KEK), Tsukuba 305-0801}\affiliation{SOKENDAI (The Graduate University for Advanced Studies), Hayama 240-0193} 
  \author{H.~Aihara}\affiliation{Department of Physics, University of Tokyo, Tokyo 113-0033} 
  \author{D.~M.~Asner}\affiliation{Pacific Northwest National Laboratory, Richland, Washington 99352} 
  \author{T.~Aushev}\affiliation{Moscow Institute of Physics and Technology, Moscow Region 141700}\affiliation{Institute for Theoretical and Experimental Physics, Moscow 117218} 
  \author{R.~Ayad}\affiliation{Department of Physics, Faculty of Science, University of Tabuk, Tabuk 71451} 
  \author{T.~Aziz}\affiliation{Tata Institute of Fundamental Research, Mumbai 400005} 
  \author{V.~Babu}\affiliation{Tata Institute of Fundamental Research, Mumbai 400005} 
% ...
\collaboration{The Belle Collaboration}


\noaffiliation

\begin{abstract}
\noindent
We measure the  decay $\Bs\to\Kz\Kzb$ using data 
collected at the $\Upsilon(5S)$ resonance with the 
Belle detector at the KEKB $e^+e^-$ collider. 
The data sample used corresponds to an integrated
luminosity of 121.4~\infb. We measure a branching
fraction 
$\mathcal{B}(\Bs\to\Kz\Kzb) =
[19.6\,^{+5.8}_{-5.1}({\rm stat.})\,\pm1.0({\rm sys.})\,\pm2.0(N^{}_{\Bs\Bsb})]\times10^{-6}$
with a significance of 5.1 standard deviations.
This measurement constitutes the first observation
of this decay.
\end{abstract}


\maketitle

\tighten

\begin{figure}[htb]
\centering
\includegraphics[width=0.45\textwidth]{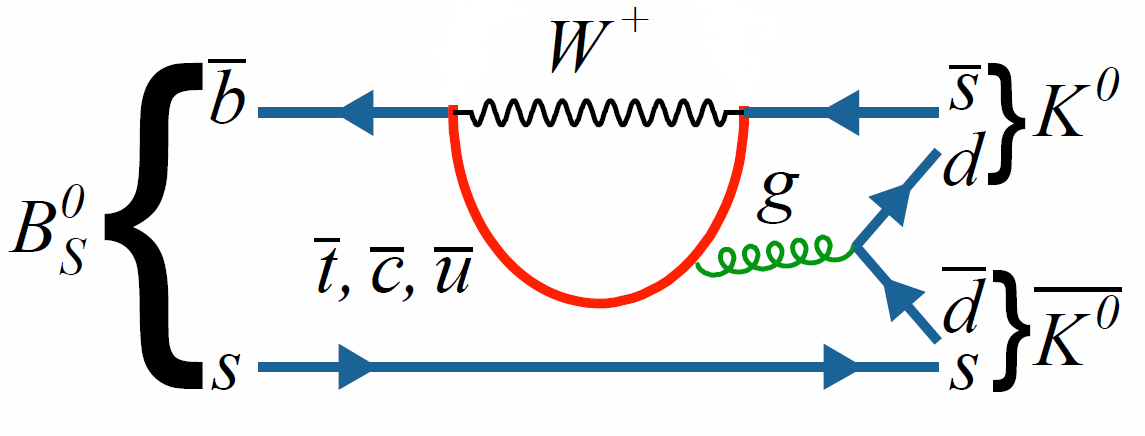}
\caption{\small Loop diagram for $\Bs\to\Kz\Kzb$ decays.  }
\label{fig:feynman}
\end{figure}

The two-body decays $\Bs\to h^+h'^-$, where $h^{\scriptscriptstyle(}\kern-1pt{}'\kern-1pt{}^{\scriptscriptstyle)}$ is
either a pion or kaon, have now all been observed~\cite{PDG}.
In contrast, the neutral-daughter decays $\Bs\to h^0h'^0$ have
yet to be observed. The decay $\Bs\to \Kz\Kzb$~\cite{charge-conjugate}
is of particular interest because the branching fraction is predicted
to be relatively large. In the standard model (SM), the decay
proceeds mainly via a $b\to s$ loop (or ``penguin") transition as shown
in Fig.~\ref{fig:feynman}, and the branching fraction is predicted
to be in the range $(16-27)\times10^{-6}$~\cite{SM-branching}.
The presence of non-SM particles or couplings could enhance 
this value~\cite{Chang:2013hba}. It has been pointed out
that $CP$ asymmetries in $\Bs\to\Kz\Kzb$ decays are
promising observables in which to search for new
physics~\cite{susy}. 

The current upper limit on the branching fraction,
$\mathcal{B}(\Bs\to\Kz\Kzb)<6.6\times 10^{-5}$ at 90\% 
confidence level, was set by the Belle Collaboration using 
$23.6~{\rm fb^{-1}}$ of data recorded at the
$\Upsilon(5S)$ resonance~\cite{Peng:2010ze}.
Here, we update this result using the full data set of
$121.4~{\rm fb^{-1}}$ recorded at the~$\Upsilon(5S)$.
The analysis presented here  uses improved tracking, $\Kz$
reconstruction, and continuum suppression algorithms.
The data set corresponds to $(6.53\pm 0.66)\times10^6$  $\Bs\Bsb$
pairs~\cite{Oswald:2015dma} produced in three $\Upsilon(5S)$ decay
channels: $\Bs\Bsb$, $\Bst\Bsb$ or $\Bs\Bstb$, and $\Bst\Bstb$.
The latter two channels dominate, with production fractions
of $f_{\Bst\Bsb}=(7.3\pm1.4)\%$ and $f_{\Bst\Bstb}=(87.0\pm1.7)$\%~\cite{Esen:2012yz}.

The Belle detector is a large-solid-angle magnetic spectrometer
consisting of a  silicon vertex detector (SVD), a 50-layer central
drift chamber (CDC), an array of aerogel threshold Cherenkov counters,
a barrel-like arrangement of time-of-flight scintillation counters,
and an electromagnetic calorimeter  comprising  CsI(Tl) crystals. 
These detector components are located inside a superconducting solenoid
coil that provides a 1.5~T magnetic field.  An iron flux-return located
outside the coil  is instrumented to detect $K_L^0$ mesons and to
identify muons. The detector is described in detail elsewhere~\cite{Belle,svd2}.
The origin of the coordinate system is defined as the position of
the nominal interaction point (IP). The $+z$ axis is aligned with
the direction opposite the $e^+$ beam
and is parallel to the direction of the magnetic field within
the solenoid. The $+x$ axis is horizontal and points towards
the outside of the storage ring; the $+y$ axis points vertically
upward.

Candidate $\Kz$ mesons are reconstructed via the decay $\KS\to\pip\pim$
using a neural network (NN) technique~\cite{Feindt:2006pm}. The NN uses
the following information:
the $\KS$ momentum in the laboratory frame;
the distance along $z$ between the two 
track helices at their closest approach;
the flight length in the $x$-$y$ plane;
the angle between the $\KS$ momentum and the vector
joining the $\KS$ decay vertex to the IP;
the angle between the
pion momentum and the laboratory-frame
direction  in
the $\KS$ rest frame;
the distance-of-closest-approach in the $x$-$y$ plane  between the IP and
the two pion helices; 
and
the pion hit information in the SVD and CDC. 
The selection efficiency is 87\% over the momentum range of interest.
We also require that the $\pip\pim$ invariant mass be
within 12~\mevm\ (about 3.5$\sigma$ in resolution)  of the nominal $\KS$ mass~\cite{PDG}.

To identify $\Bs\to\KS\KS$ candidates, we define two
variables: the beam-energy-constrained mass
$M_{\rm bc}=\sqrt{E^2_{\rm beam}-|\vec{p}^{}_{B}|^2c^2}/c^2$;
and the energy difference 
$\Delta E=E_{B}-E_{\rm beam}$, where $E_{\rm beam}$ is the beam
energy and $E_B$ and $\vec{p}^{}_{B}$ are the energy and momentum,
respectively, of the $\Bs$ candidate. These quantities are
evaluated in the $e^+e^-$ center-of-mass  frame.
We require that events satisfy $M_{\rm bc} > 5.34$~\gevm\ 
and $-0.20~{\rm GeV} < \Delta E< 0.10~{\rm GeV}$.

To suppress background arising from continuum $e^+e^-\to q\qbar~(q=u,d,s,c)$
production, we use a second NN~\cite{Feindt:2006pm} that distinguishes jetlike continuum
events from more spherical $\Bstar\Bstarb$ events.
This NN uses the following input  variables, which characterize the event topology:
the cosine of the angle between the thrust axis~\cite{Brandt:1964sa}
of the $\Bs$ candidate and the thrust axis of the rest of the event;
the cosine of the angle between the $\Bs$  thrust axis and the $+z$
axis; a set of 16 modified Fox-Wolfram moments~\cite{SFW}; and 
the ratio of the second to zeroth (unmodified) Fox-Wolfram moments.
All quantities are evaluated in the  center-of-mass frame.
The NN is trained using Monte Carlo (MC) simulated signal events 
and $q\qbar$ background events. The MC samples are obtained using
\evtgen~\cite{Lange:2001uf} for event generation and \geant~\cite{geant3}
for modeling the detector response. The NN has a single output variable
($C_{\rm NN}$) that ranges from $-1$ for backgroundlike events to $+1$
for signal-like events. We require $C_{\rm NN}>-0.1$, which rejects 
approximately 85\% of $q\qbar$ background while retaining 83\%
of signal decays. We subsequently translate $C_{\rm NN}$ to a new
variable 
\begin{linenomath}
\begin{equation}
C'_{\rm NN} = \ln\left(\frac{C_{\rm NN}-C^{\rm min}_{\rm NN}}
{C^{\rm max}_{\rm NN}-C_{\rm NN}}\right),
\end{equation} 
\end{linenomath}
where $C^{\rm min}_{\rm NN}= -0.1$ and $C_{\rm NN}^{\rm max}$
is the maximum value of $C_{\rm NN}$ obtained from a large
sample of signal MC decays.
The distribution of $C'_{\rm NN}$ is  well modeled by a Gaussian
function.

After applying all selection criteria, approximately 1.0\% 
of events have multiple $\Bs$ candidates. For these
events, we retain the candidate having the
smallest value  of $\chi^2$ obtained from 
the deviations of the reconstructed $\KS$  masses from their nominal values~\cite{PDG}.
According to MC simulation, this criterion selects the correct $\Bs$
candidate $>99$\% of the time.

We measure the signal yield by performing an unbinned extended
maximum likelihood fit to the variables $M_{\rm bc}$, $\Delta E$,
and $C^{\prime}_{\rm NN}$. The likelihood function is defined as
\begin{linenomath}
\begin{equation}
\mathcal{L} = e^{-\sum_j Y_j}
\prod_i^N \left( \sum_j Y_j \mathcal{P}_j(M_{\rm bc}^i, \Delta E^i, C'^i_{\rm NN} )\right),
\end{equation} 
\end{linenomath}
where $Y_j$ is the yield of  component~$j$;
$\mathcal{P}_j(M_{\rm bc}^i, \Delta E^i, C'^i_{\rm NN})$ is the
probability density function (PDF) 
 of component $j$ for event $i$;
$j$ runs over the two event categories (signal and $q\qbar$ background);
and $i$ runs over all events in the sample~($N$).
Backgrounds arising from other $\Bs$ and non-$\Bs$ decays
were studied using MC simulation and found to be negligible. 
As correlations among the variables $M_{\rm bc}$, $\Delta E$,
and $C^{\prime}_{\rm NN}$ are found to be small, the three-dimensional PDFs
$\mathcal{P}_j(M_{\rm bc}^i, \Delta E^i, C'^i_{\rm NN})$ are factorized 
into the product of separate one-dimensional PDFs.

The signal PDF is defined as
\begin{linenomath}
\begin{eqnarray}
\mathcal{P}_{\rm sig} & = & 
f_{\Bst\Bstb}\mathcal{P}_{\Bst\Bstb}+f_{\Bst\Bsb}\mathcal{P}_{\Bst\Bsb} \\ \nonumber
 & & \hskip0.20in + (1-f_{\Bst\Bstb}-f_{\Bst\Bsb})\mathcal{P}_{\Bs\Bsb},
\end{eqnarray} 
\end{linenomath}
where 
$\mathcal{P}_{\Bst\Bstb}$, $\mathcal{P}_{\Bst\Bsb}$, and 
$\mathcal{P}_{\Bs\Bsb}$ are the PDFs for signal arising from
$\Upsilon(5S)\to\Bst\Bstb,\,(\Bst\Bsb+\Bstb\Bs)$, and 
$\Bs\Bsb$ decays.
The $M_{\rm bc}$ and $C'_{\rm NN}$ PDFs are modeled  with
Gaussian functions, and the $\Delta E$ PDFs are each modeled
with a sum of two Gaussian functions having a common mean. 
All parameters of the  signal PDF are fixed to the corresponding MC  values.
The peak positions for $M_{\rm bc}$ and $\Delta E$ are adjusted according
to small data-MC differences observed in a control sample of 
$\Bs\to D_s^- \pi^+$ decays~\cite{Esen:2012yz}.
As this control sample has only modest statistics,
the resolutions for $M_{\rm bc}$,
$\Delta E$, and $C^{\prime}_{\rm NN}$, and the peak position
for $C^{\prime}_{\rm NN}$, are adjusted for data-MC differences using a high statistics sample of
$B^0\to D^-(\to  K^+\pi^-\pi^-)\pi^+ $ decays. For $q\qbar$
background, the $M_{\rm bc}$, $\Delta E$, and $C^{\prime}_{\rm NN}$
PDFs are modeled with an ARGUS function~\cite{Albrecht:1990am},
a first-order Chebyshev polynomial, and a Gaussian function,
respectively. All parameters of the $q\qbar$ background PDFs 
except for the end point of the ARGUS function are floated in
the fit.

\begin{figure*}[htb]
  \includegraphics[width=0.32\textwidth]{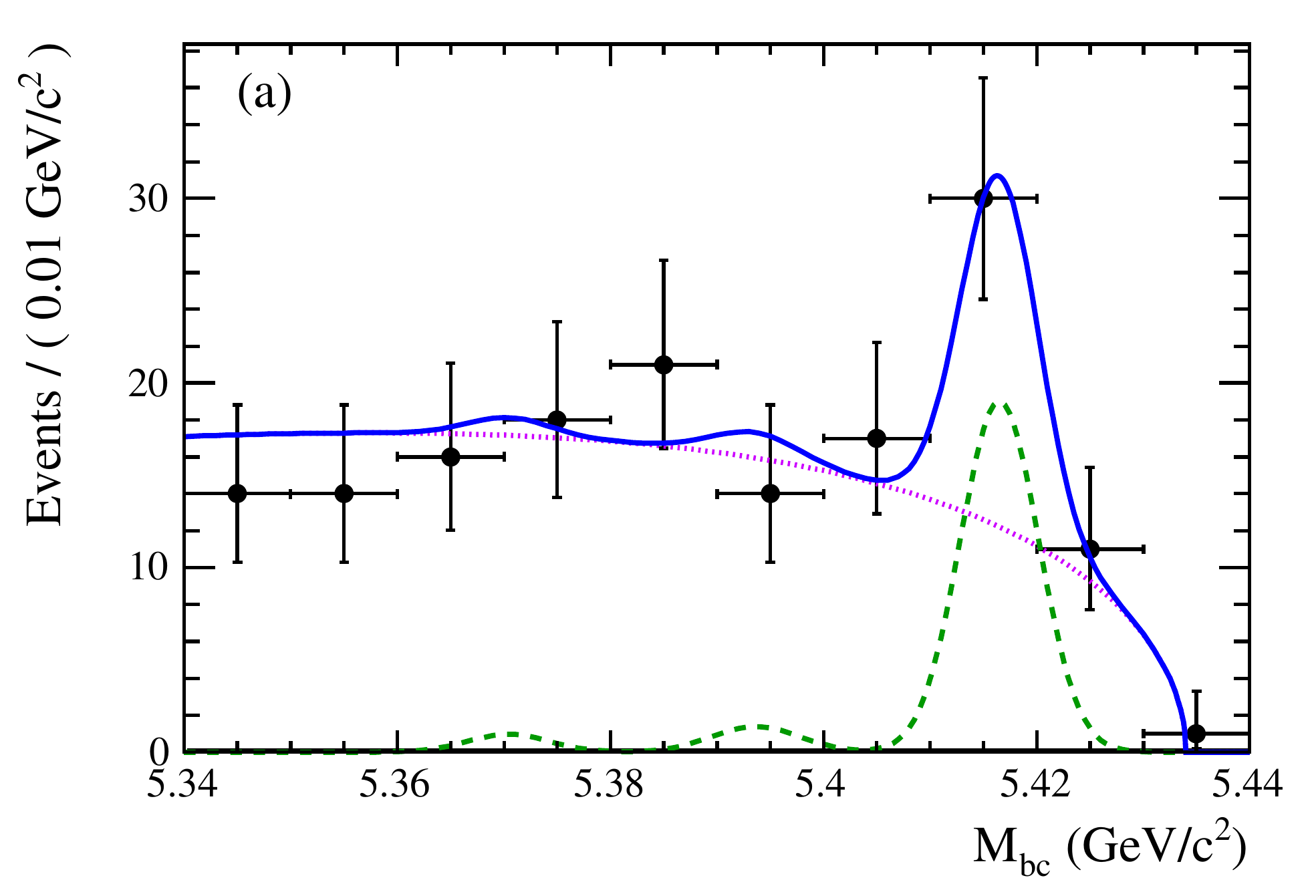}
  \includegraphics[width=0.32\textwidth]{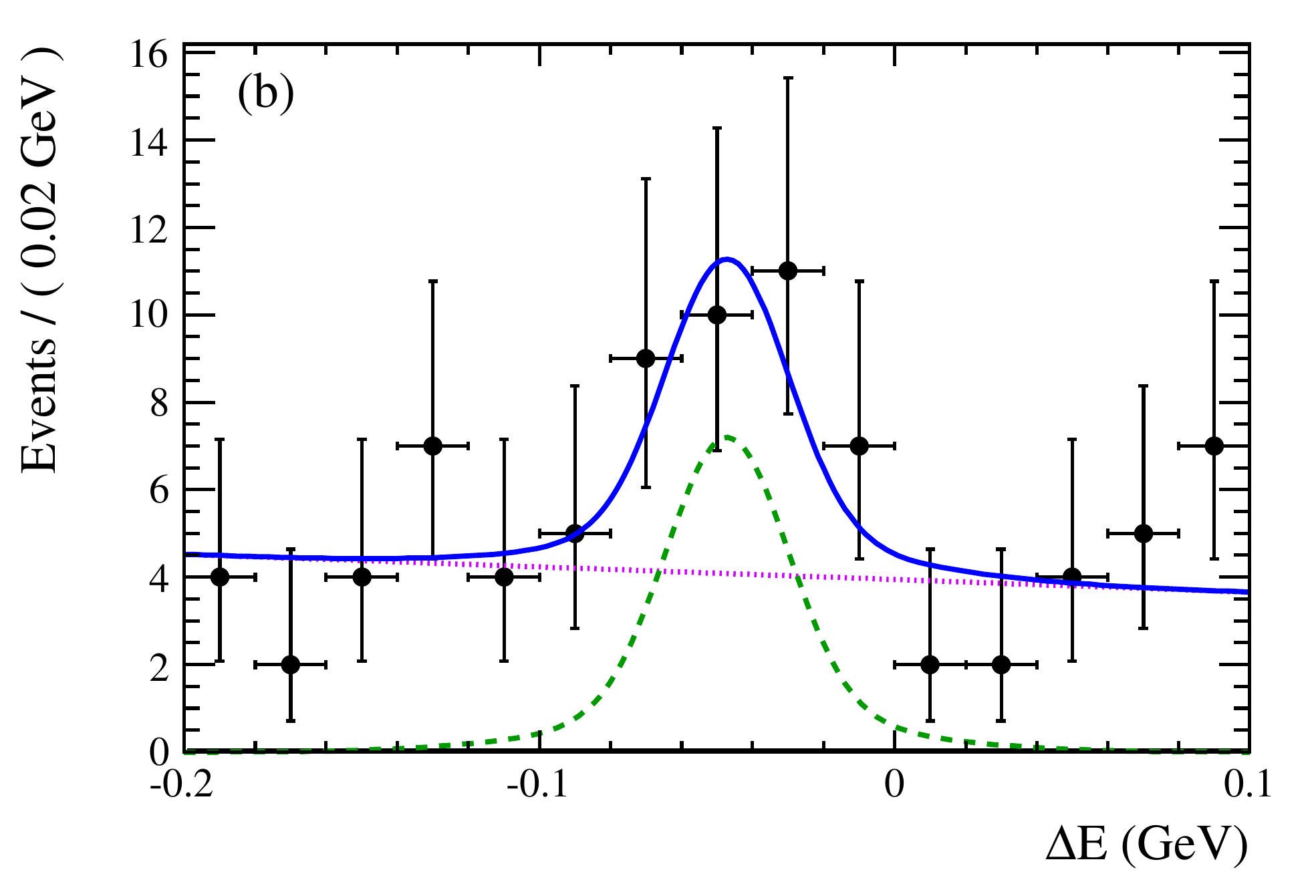}
  \includegraphics[width=0.32\textwidth]{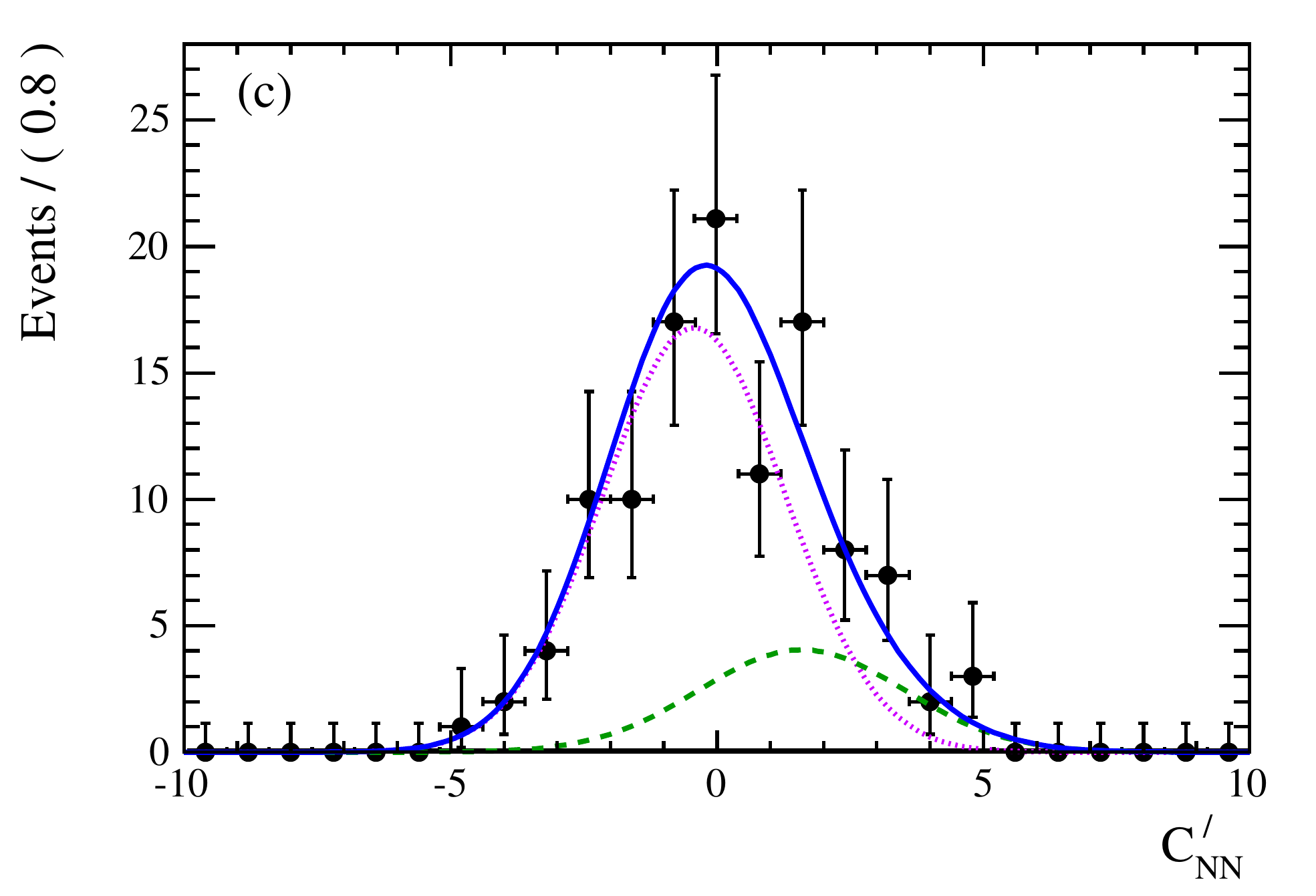}
\caption{\small  Projections of the 3D fit to the real data: 
(a) $M_{\rm bc}$ in $-0.11~{\rm GeV} <\Delta E < 0.02~{\rm GeV}$
and $C^{\prime}_{\rm NN}>0.5$;
(b) $\Delta E$ in $5.405~{\rm GeV}/c^{2} <M_{\rm bc}< 5.427~{\rm GeV}/c^{2}$
and $C^{\prime}_{\rm NN}>0.5$; and 
(c) $C^{\prime}_{\rm NN}$ in $5.405~{\rm GeV}/c^{2} <M_{\rm bc}< 5.427~{\rm GeV}/c^{2}$
and $-0.11~{\rm GeV} <\Delta E < 0.02~{\rm GeV}$. 
The points with  error bars are data, the (green) dashed curves
show the signal, (magenta) dotted curves show the continuum
background, and (blue) solid  curves show the total. The $\chi^2/{\rm (number~ of~ bins)}$ values of these fit projections are 0.30, 0.43, and 0.26, respectively, which indicate that the fit gives a good description of the data. The three peaks in $M_{\rm bc}$ arise from 
$\Upsilon(5S)\to\Bs\Bsb, \Bst\Bsb+\Bs\Bstb$, and $\Bst\Bstb$ decays.}
\label{fig:fig2}
\end{figure*}

The results of the fit are $29.0\,^{+8.5}_{-7.6}$ signal events
and $1095.0\,^{+33.9}_{-33.4}$ continuum background events.
Projections of the fit are shown in Fig.~\ref{fig:fig2}.
The branching fraction is calculated via
\begin{linenomath}
\begin{eqnarray}
\mathcal{B}(\Bs\to\Kz\Kzb) & = &
\frac{Y_{s}}{2 N_{\Bs\Bsb}(0.50)
\mathcal{B}^2_{\Kz} \varepsilon},
\end{eqnarray}
\end{linenomath}
where $Y_{s}$ is the fitted signal yield;
$N_{\Bs\Bsb}=(6.53\pm 0.66)\times10^6$ is the number of $\Bs\Bsb$ events;
$\mathcal{B}_{\Kz}=(69.20\pm0.05)\%$ is the branching fraction for $\KS\to\pi^+\pi^-$~\cite{PDG};
and 
$\varepsilon=(46.3\pm 0.1)\%$ is the signal efficiency as determined from MC simulation. The efficiency is corrected by a factor 
$1.01\pm0.02$ for each reconstructed $\KS$, to account for a small difference in $\KS$ reconstruction efficiency between data and simulation. 
This correction is estimated from a high statistics sample of $D^0\to\KS\pi^0$ decays.
The factor 0.50 accounts for the 50\% probability for 
$\Kz\Kzb\to\KS\KS$ (since $\Kz\Kzb$ is $CP$ even).
Inserting these values gives
$\mathcal{B}(\Bs\to\Kz\Kzb)=(19.6\,^{+5.8}_{-5.1})\times10^{-6}$,
where the error is statistical.

The systematic uncertainty on $\mathcal{B}(\Bs\to\Kz\Kzb)$
arises from several sources, as listed in Table~\ref{sys1}. 
The uncertainties due to the fixed parameters in the PDF shape 
are estimated by varying the parameters individually according
to their statistical uncertainties. For each variation the
branching fraction is recalculated, and the difference with
the nominal branching fraction is taken as the systematic
uncertainty associated with that parameter. We add together
all uncertainties in quadrature to obtain the overall
uncertainty due to fixed parameters.
The uncertainties due to errors in the calibration factors and
the fractions $f^{}_{B^{(*)}_s \overline{B}^{\,(*)}_s}$ are evaluated
in a similar manner. 
To test the stability of our fitting procedure, we generate and fit a large ensemble of MC
pseudoexperiments. By comparing the mean of the fitted yields with the input value, a  bias of $-2.6\%$ is found. We attribute this bias to our neglecting small correlations among the fitted observables.
An 0.9\% systematic uncertainty is assigned due to the $C_{\rm NN}$
selection; this is obtained by comparing the selection efficiencies
in MC simulationand data for the $B^0\to D^-(\to K^{+}\pim\pim)\pip$ control
sample. We assign a 2.0\% systematic uncertainty for each
reconstructed $\KS\to\pip\pim$; this is determined using 
a $D^0\to\KS\pi^0$ sample. The uncertainty
on $\varepsilon$ due to the MC sample size is 0.2\%. 
The total of the above systematic uncertainties is calculated as their sum in quadrature. In addition, there is
a 10.1\% uncertainty due to the number of $\Bs\Bsb$ pairs. As this large uncertainty does not arise from our analysis, we quote it separately.

\begin{table}[htb]
\renewcommand{\arraystretch}{1.2}
\caption{\small Systematic uncertainties on $\mathcal{B}(\Bs\to\Kz\Kzb)$.
Those listed in the upper section are associated with fitting for the 
signal yields and are included in the signal significance. }
\label{sys1}
\centering
\begin{tabular}{l|c}
\hline \hline
Source & Uncertainty (\%) \\
\hline
PDF parametrization & 0.2\\
Calibration factor &$^{+0.9}_{-0.8}$\\
$f^{}_{B^{(*)}_s \overline{B}^{\,(*)}_s}$ & $^{+1.2}_{-1.1}$\\
Fit bias            & $^{+0.0}_{-2.6}$\\
\hline
$\KS\to\pip\pim$  reconstruction & 4.0 \\
$C_{\rm NN}$ selection & 0.9 \\ 
MC sample size & 0.2\\
$\mathcal{B}($\KS\to\pip\pim$)$ & 0.1 \\
\hline
Total (without $N^{}_{\Bs\Bsb}$) & $^{+4.4}_{-5.1}$ \\
$N^{}_{\Bs\Bsb}$ & 10.1\\
\hline\hline
\end{tabular}
\end{table}

The signal significance is calculated as
$\sqrt{-2\ln(\mathcal{L}_0/\mathcal{L}_{\rm max})}$, where
$\mathcal{L}_0$  is the likelihood value when the signal
yield is fixed to zero, and $\mathcal{L}_{\rm max}$ is the 
likelihood value of the nominal fit. We include systematic
uncertainties in the significance by convolving the likelihood
function with a Gaussian function whose width is equal to
that part of the systematic uncertainty that affects the
signal yield. We obtain a signal
significance of 5.1 standard deviations;  thus, our measurement constitutes
the first observation of this decay. 



\begin{figure}[htb]
  \includegraphics[width=0.38\textwidth]{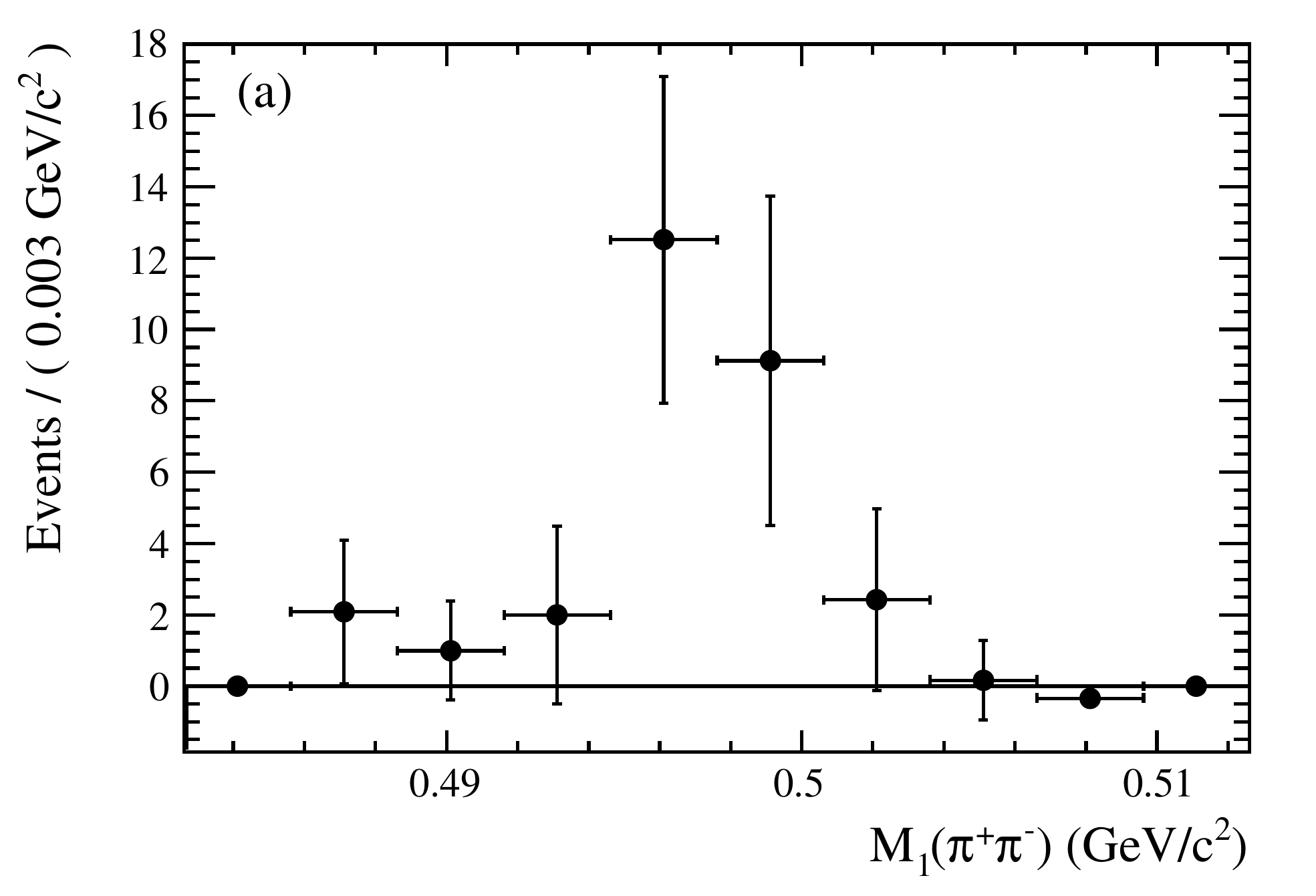}
  \includegraphics[width=0.38\textwidth]{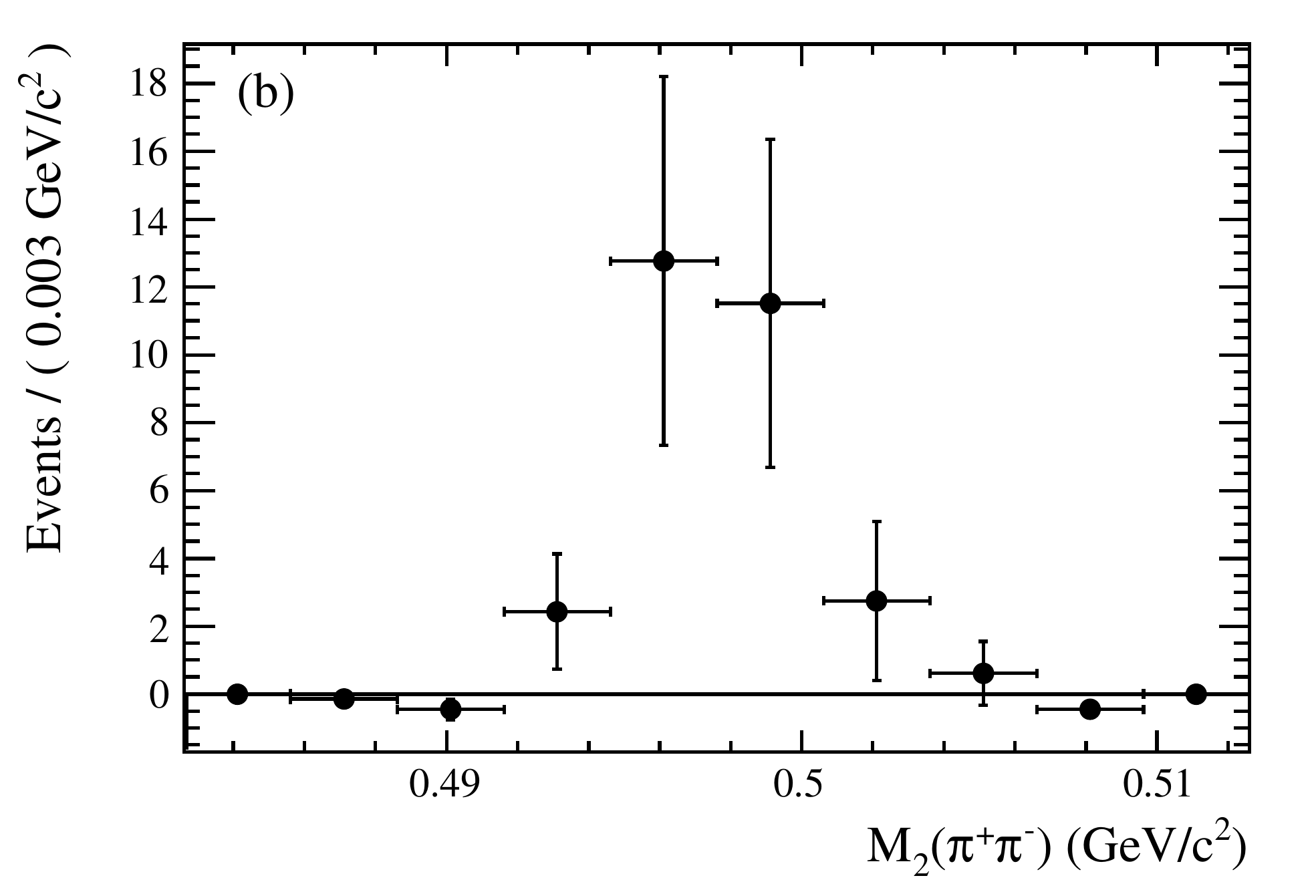}
\caption{\small  Background subtracted {\it sPlot} distributions of $M(\pip\pim)$ for the   (a) higher momentum and (b)  lower momentum  $\KS$  candidates.}
\label{fig:splot}
\end{figure}

Figure~\ref{fig:splot} shows the  background-subtracted {\it sPlot}~\cite{splot} distributions of $M(\pip\pim)$, where the $\KS$ selection is removed for the $\pip\pim$ pair being plotted. No $\Bs\to\KS\pip\pim$ contribution is observed. We check this quantitatively by performing our signal fit for events  in the mass sidebands of each $\KS$  [$M(\pip\pim)\in (0.460,0.485)~{\rm GeV}/c^2$ and  $M(\pip\pim)\in (0.510,0.530)~{\rm GeV}/c^2$]. The extracted signal yields,  
$-0.7\,^{+2.9}_{-2.1}$ and $1.6\,^{+2.2}_{-1.2}$  for the higher momentum $\KS$ and lower momentum $\KS$, respectively, are consistent with zero.  We calculate the expected number of  $\Bs\to\KS\pip\pim$ events in our signal sample using MC simulation and the measured branching fraction, $\mathcal{B}( \Bs\to K^0\pip\pim)=15.0\times10^{-6}$~\cite{Aaij:2013uta}; the result is 0.001.

In summary, we report the first observation of the decay
$\Bs\to\Kz\Kzb$. The branching fraction is measured to be 
\begin{linenomath}
\begin{equation*}
\mathcal{B}(\Bs\to\Kz\Kzb)=(19.6\,^{+5.8}_{-5.1}\,\pm1.0\,\pm2.0)\times10^{-6},
\end{equation*}
\end{linenomath} 
where the first uncertainty is statistical, the second
is systematic, and the third reflects the uncertainty due to the 
total number of $\Bs\Bsb$ pairs. This value  is in good agreement with the
SM predictions~\cite{SM-branching}, and it implies that the Belle II experiment~\cite{Abe:2010gxa} will
reconstruct over 1000 of these decays. Such a sample would allow for a much higher sensitivity search for new physics in this $b\to s$ penguin-dominated decay.

\begin{center}
\textbf{ACKNOWLEDGMENTS}
\end{center}
We thank the KEKB group for the excellent operation of the
accelerator; the KEK cryogenics group for the efficient
operation of the solenoid; and the KEK computer group,
the National Institute of Informatics, and the 
PNNL/EMSL computing group for valuable computing
and SINET4 network support.  We acknowledge support from
the Ministry of Education, Culture, Sports, Science, and
Technology (MEXT) of Japan, the Japan Society for the 
Promotion of Science (JSPS), and the Tau-Lepton Physics 
Research Center of Nagoya University; 
the Australian Research Council;
Austrian Science Fund under Grants No.~P 22742-N16 and P 26794-N20;
the National Natural Science Foundation of China under Contracts 
No.~10575109, No.~10775142, No.~10875115, No.~11175187, and  No.~11475187;
the Chinese Academy of Science Center for Excellence in Particle Physics; 
the Ministry of Education, Youth and Sports of the Czech
Republic under Contract No.~LG14034;
the Carl Zeiss Foundation, the Deutsche Forschungsgemeinschaft
and the VolkswagenStiftung;
the Department of Science and Technology of India; 
the Istituto Nazionale di Fisica Nucleare of Italy; 
the WCU program of the Ministry of Education, National Research Foundation (NRF) 
of Korea Grants No.~2011-0029457,  No.~2012-0008143,  
No.~2012R1A1A2008330, No.~2013R1A1A3007772, No.~2014R1A2A2A01005286, 
No.~2014R1A2A2A01002734, No.~2015R1A2A2A01003280 , No. 2015H1A2A1033649;
the Basic Research Lab program under NRF Grant No.~KRF-2011-0020333,
Center for Korean J-PARC Users, No.~NRF-2013K1A3A7A06056592; 
the Brain Korea 21-Plus program and Radiation Science Research Institute;
the Polish Ministry of Science and Higher Education and 
the National Science Center;
the Ministry of Education and Science of the Russian Federation and
the Russian Foundation for Basic Research;
the Slovenian Research Agency;
the Basque Foundation for Science (IKERBASQUE) and 
the Euskal Herriko Unibertsitatea (UPV/EHU) under program UFI 11/55 (Spain);
the Swiss National Science Foundation; the National Science Council
and the Ministry of Education of Taiwan; and the U.S.\
Department of Energy and the National Science Foundation.
This work is supported by a Grant-in-Aid from MEXT for 
Science Research in a Priority Area (``New Development of 
Flavor Physics'') and from JSPS for Creative Scientific 
Research (``Evolution of Tau-lepton Physics'').


\begin{thebibliography}{99}
\bibitem{charge-conjugate}
Unless stated otherwise, charge-conjugate modes are implicitly included.
\end{thebibliography}
\end{document}